\documentclass[11pt,a4paper]{article}

\usepackage{./jheppub}
\usepackage{color}

\def\dfrac#1#2{\displaystyle\frac{#1}{#2}}

\newcommand{\p}{\partial}
\newcommand{\pslash}{p\kern-1ex /}
\newcommand{\qslash}{q\kern-1ex /}
\newcommand{\lslash}{l\kern-1ex /}
\newcommand{\sslash}{s\kern-1ex /}
\newcommand{\kaslash}{k_a\kern-2ex /}
\newcommand{\kbslash}{k_b\kern-2ex /}
\newcommand{\Dslash}{{\cal D}\kern-1.5ex /}

\newcommand{\beqa}{\begin{eqnarray}}
\newcommand{\eeqa}{\end{eqnarray}}

\begin{document}
\title{Is energy conserved in general relativity ?}
\author[a]{Sinya AOKI}
\affiliation[a]{Fundamental Quantum Science Program, TRIP Headquarters, RIKEN, Hirosawa 2-1, 
Wako, Saitama 351-0198, Japan}
\affiliation[b]{Center for Gravitational Physics and Quantum Information, Yukawa Institute for Theoretical Physics, Kyoto University, Kitashirakawa Oiwakecho, Sakyo-ku, Kyoto 606-8502, Japan}
\abstract{This short report  is dedicated to the 40th anniversary of International Journal of Modern Physics A (IJMPA) and Modern Physics Letters A (MPLA). 
While  the report  is based on a series of papers\cite{Aoki:2020prb,Aoki:2020nzm,Aoki:2022gez,Aoki:2022ugd,Aoki:2022ysm,Aoki:2022tek,Aoki:2023zoq,Aoki:2023ufz},
its content reflects my personal viewpoints. Therefore I am solely responsible for all the statements in the report.
In this report we discuss conservation of energies in a curved spacetime including general relativity. 
We argue that the matter energy is not necessarily conserved in a curved space time due to a lack of time translational invariance, and
adding energy of gravitational fields to recover the conservation law of energies fails due to Noether's 2nd theorem.
We show that there exists  conserved quantities associated with the matter energy momentum tensor other than the energy in a curved spacetime,
and discuss consequences of the existence of such conserved charges. 
  }
\preprint{FQSP-25-5, YITP-25-196}
\maketitle

\section{Introduction}
A few yeas ago, I wrote an essay\cite{Aoki:2022mlc},  dedicated to Prof. K.K. Phua on the occasion of his 80th birthday,
entitled  ``Do we know how to define energy in general relativity ?''.
In this report, I consider the same problem in a slightly different point of view.

Why do we think that  the ``energy'' is conserved in a physical system ?
A simple answer to this question is as follows.
An  invariance under time translation leads to a conserved energy as the Noether charge\cite{Noether:1918zz}.
A generic translation is given by $x^\mu\to x^\mu + a\xi^\mu$, where 
$a$ and $\xi^\mu$ are constant scalar and vector, respectively, and
$x=(x^0,x^1,x^2,x^3)$ is a point in the spacetime
with time coordinate $x^0$ and space coordinates $x^k$ ($k=1,2,3$).   
The time translation is generated by a vector $\xi^\mu_T= -\delta^\mu_0$ as $x^0\to x^0-a$ and $x^k\to x^k$.

If a system has an explicit $x^0$ dependence which violates the invariance under the time-translation such as an external electric field, there is no reason for the corresponding energy to conserve. 
Thus the energy conservation is non-trivial in general relativity\cite{Einstein:1916}, since the curved spacetime can become time-dependent. 

\section{Energy (non-)conservation in a curved space time}
A curved spacetime is characterized by a metric $g_{\mu\nu}$, which defined a ``distance''  $ds$ between $x^\mu$ and $x^\mu+dx^\mu$ as
\beqa
ds^2= g_{\mu\nu}(x) dx^\mu dx^\nu.
\eeqa
For example, the flat ({\it i.e.}, non-curved) space-time is described by $g_{\mu\nu} :={\rm diag}(-1,1,1,1)$.

The energy, defined as  a naive extension of the conserved charge in the flat spacetime for the time translation, becomes
\beqa
E(x^0) &=& \int_{\Sigma_{x^0}} d\Sigma_\mu T^\mu{}_\nu \xi^\nu_T = -\int_{x^0={\rm const.}} d^3x\, \sqrt{-g} T^0{}_0,
\label{eq:def_energy}
\eeqa
 where $T^\mu{}_\nu$ is the energy-momentum tensor (EMT) in the system, $\Sigma_{x^0}$ is a hypersurface defined by the constant $x^0$
 with its surface element $d\Sigma_\mu$,  and $g=\det g_{\mu\nu}$. In the last equality, we take $x^k$ as the space coordinate on the hypersurface.  

If $g_{\mu\nu}(x)$ depends on the time coordinate $x^0$, the energy defined above may not be conserved due to a lack of translational invariance.  

\subsection{Energy non-conservation in general relativity: Expanding Universe}
Let us consider a well-know example of the spacetime in general relativity, which manifestly shows energy non-conservation.
That is a homogeneous and isotropic  expanding Universe, described by Friedmann-Lemaitre-Robertson-Walker metric\cite{Friedmann:1924bb, Lemaitre:1931zza,Robertson:1935zz,Walker:1937aa} as
\beqa
ds^2 = -(dx^0)^2 + a^2(x^0) \tilde g_{ij} dx^i dx^j, 
\eeqa 
where $a(x^0)$ controls a size of Universe, and we take the space part as the 3-sphere: $\bar g_{ij} dx^idx^j = d\psi^2 +\sin^2\psi (d\theta^2+\sin^2\theta d\phi^2)$ with $0\le \psi,\theta\le \pi$ and $0\le \phi \le 2\pi$, which represents the close Universe.

The corresponding EMT is given by  the perfect fluid
\beqa
T^0{}_0 = -\rho(x^0), \quad T^i{}_j =P(x^0) \delta^i_j \ (i,j =1,2,3),
\eeqa
where $\rho(x^0)$ is the energy density and $P(x^0)$ is the pressure, and they depend only on time $x^0$.
The (covariant) conservation  of the EMT $\nabla_\mu T^\mu{}_\nu =0$ implies
\beqa
\dot\rho(x^0) + 3\left[\rho(x^0) + P(x^0)\right] H(x^0) = 0,\quad H(x^0):= {\dot a(x^0)\over a(x^0)},
\eeqa 
where $\dot{f}$ means the derivative with respect to $x^0$.

According to the definition \eqref{eq:def_energy}, the energy becomes
\beqa
E(x^0) = \int d^3x \sqrt{\tilde g} a^3(x^0) \rho(x^0) = V_3 a^3(x^0) \rho(x^0),
\eeqa
where we use $\sqrt{-g}= a^3(x^0)\sqrt{\tilde g}$, and the constant space volume factor $V_3$ is given by
\beqa
V_3 = \int d^3 x \sqrt{\tilde g} = \int d\psi\, d\theta\, d\phi\, \sin^2\psi\,\sin\theta={2\pi^2}.
\eeqa
Thus its time-derivative is given by
\beqa
\dot E(x^0) = V_3 a^3(x^0)\left[ \dot\rho(x^0) + 3 H(x^0) \rho(x^0)\right]
= - 3 V_3 a^2(x^0) \dot a(x^0) P(x^0),
\eeqa
which is non-zero in general if $P(x^0)\not=0$. The energy decreases or increases since the spacetime expand against $P>0$ or $P<0$, respectively.
The matter performs positive or negative work to the spacetime ``piston''.
In this example, clearly the matter energy is not conserved.
Where has the energy used by work gone ? We consider this question later.

\subsection{Energy conservation in a curved spacetime}
We next consider opposite cases, where the energy is conserved in a curved spacetime.

\subsubsection{Conservation with symmetry}
The first one is rather trivial but still instructive.
If the vector $\xi^\mu_T$ for the time translation is the Killing vector such that
\beqa
\nabla_\mu (\xi_T)_\nu + \nabla_\nu(\xi_T)_\mu = 0,
\eeqa   
then $J^\mu := T^\mu{}_\nu \xi_T^\nu$ is covariantly conserved as\cite{Aoki:2020prb}
\beqa
\nabla_\mu (T^\mu{}_\nu \xi_T^\nu) = T^{\mu\nu} \nabla_\mu(\xi_T)_\nu ={1\over 2} T^{\mu\nu}\left[ \nabla_\mu (\xi_T)_\nu + \nabla_\nu(\xi_T)_\mu \right]= 0.
\eeqa
In this case,  the time translation generated by $\xi^\mu_T$ becomes a symmetry of the system  and therefore
$J^\mu$ is the Noether current associated with the symmetry, and the energy $E(x^0)$ is the conserved Noether charge.
It is also shown that $\xi^\mu_T$ becomes the Killing vector if the metric $g_{\mu\nu}$ doesn't depend on $x^0$.

As a concrete example, we consider a spherically symmetric compact star, where the metric is given by
\beqa
ds^2 = - f(r) (dx^0)^2 + h(r) dr^2 + r^2 d\Omega^2, \quad d\Omega^2 =d\theta^2 +\sin^2\theta d\phi^2,
\eeqa  
and the corresponding EMT is the perfect fluid as $T^0{}_0=-\rho(r)$ and $T^i{}_j = P(r)\delta^i_j$\cite{Aoki:2020prb}.
Outside the star ($ r> R$), the EMT vanishes and  the metric describes the Schwarzschild vacuum as 
\beqa
f(r) ={1\over h(r)} = 1 -  {2G_N M(R)\over r}, 
\eeqa
where $M(R)$ is called the Misner-Sharp mass\cite{Misner:1964je} and given by
\beqa
M(R) = \int d^3x\, \rho(r) = 4\pi \int_0^R dr\, r^2 \rho(r),
\label{eq:MS_mass}
\eeqa
and $G_N$ is the Newton constant.

While the Misner-Sharp mass $M(R)$ is {\it NOT} covariant,
the energy is  covariant by definition and given by
\beqa
E = \int d^3x\, \sqrt{-g} T^0{}_\nu \xi_T^\nu =4\pi \int_0^R dr\, r^2\, \sqrt{f(r)h(r)} \rho(r), 
\eeqa
which is clearly different from $M(R)$.
This means that the energy for the matter of which the compact star is composed is different from the gravitational mass in the metric, 
which is a type of the quasilocal energy in general relativity such as the ADM mass\cite{Arnowitt:1962hi}.

What causes the difference between them ? 
We rewrite the difference as
\beqa
\Delta E := E - M(R) = - 4\pi G_N\int_0^R dr\, \sqrt{f(r) h^3(r)} r M(r) \left\{ \rho(r) + P(r)\right\},
\eeqa
where 
\beqa
M(r) = 4\pi \int_0^r ds\, s^2 \rho(s).
\label{eq:MS_mass2}
\eeqa
In the Newtonian limit, we have
\beqa
\Delta E  \simeq -4\pi G_N \int_0^\infty dr r\, M(r) \rho(r),
\eeqa
while the gravitational potential energy in Newtonian dynamics becomes
\beqa
U := -{G_N\over 2} \int d^3x\, d^3y {\rho(\vert{\bf x}\vert)\rho(\vert{\bf y}\vert)\over \vert {\bf x} -{\bf y}\vert}
= -4\pi G_N \int_0^\infty dr r\, M(r) \rho(r).
\eeqa
We thus obtain $\Delta E = U$ in this limit, so that $\Delta E < 0$.

Thus the total energy including the gravitational potential energy, which is a familiar one in the flat spacetime, is  less than  the gravitational mass seen far from the star. In the standard textbooks, $M(R)$ is regarded as  the ``energy'' of the compact star, 
but interpretations on the difference between $E$ and $M(R)$ disagree among them\cite{Misner:1974qy,Hawking:1973uf,Wald:1984rg,Weinberg:1972kfs,Schutz:1985jx}.
See appendix in Ref.~\cite{Aoki:2020prb} for more details.
For the constant density $\rho(r)=\rho_0$, the maximum value of $\vert \Delta E\vert$ becomes as large as 68\% of $M(R)$\cite{Aoki:2020prb}.

\subsubsection{Conservation without symmetry}
If the vector $\xi_T^\mu$ for the time translation satisfies
\beqa
T^\mu{}_\nu \nabla_\mu \xi_T^\nu = -T^\mu{}_\nu \Gamma^\nu_{\mu0} = 0,
\eeqa
the vector $J^\mu= T^\mu{}_\nu \xi_T^\nu$ is conserved as $\nabla_\mu J^\mu = T^\mu{}_\nu \nabla_\mu \xi_T^\nu =0$.
Although $J^\mu$ is not a Noether current in this case, we can define the conserved energy $E$.

As an example, we consider a model of gravitational collapse, whose metric is given in the Eddington-Finkelstein coordinate as
\beqa
ds^2 = -\left( 1+ u\right) dt^2  - 2 u dt dr + \left(1- u\right) dr^2 + r^2 d\Omega^2,
\eeqa
where we use $t$ instead of $x^0$ for the time coordinate, and $u$ is a function of $r,t$ as
\beqa
u(r,t) = - {2 m(r,t)\over r} .
\eeqa
A good property of this coordinate is that constant $t$ hypersurfaces are alway space-like even within the horizon ( $ 1+ u < 0$ ).
Thus the coordinate $t$ can be regarded as ``time".    

The time dependent mass function $m(r,t)$ is taken as\cite{Aoki:2024dyr}
\beqa
m(r,t) =
\left\{
\begin{array}{ll}
    h \left( x(t)-x^3(t) + {x^4(t)\over 2}\right), & x(t) := \dfrac{r}{he^{-\omega t}} \le 1,   \\
   \\
    {\cal M}:=\dfrac{h}{2} & x(t) > 1,     \\
\end{array}
\right.
\label{eq:mrt}
\eeqa
where $h$ the horizon size of the would-be black hole (BH), and thus ${\cal M}=h/2$ is the mass of the would-be BH.

\begin{figure}
	\centering
	\includegraphics[width=0.50\textwidth]{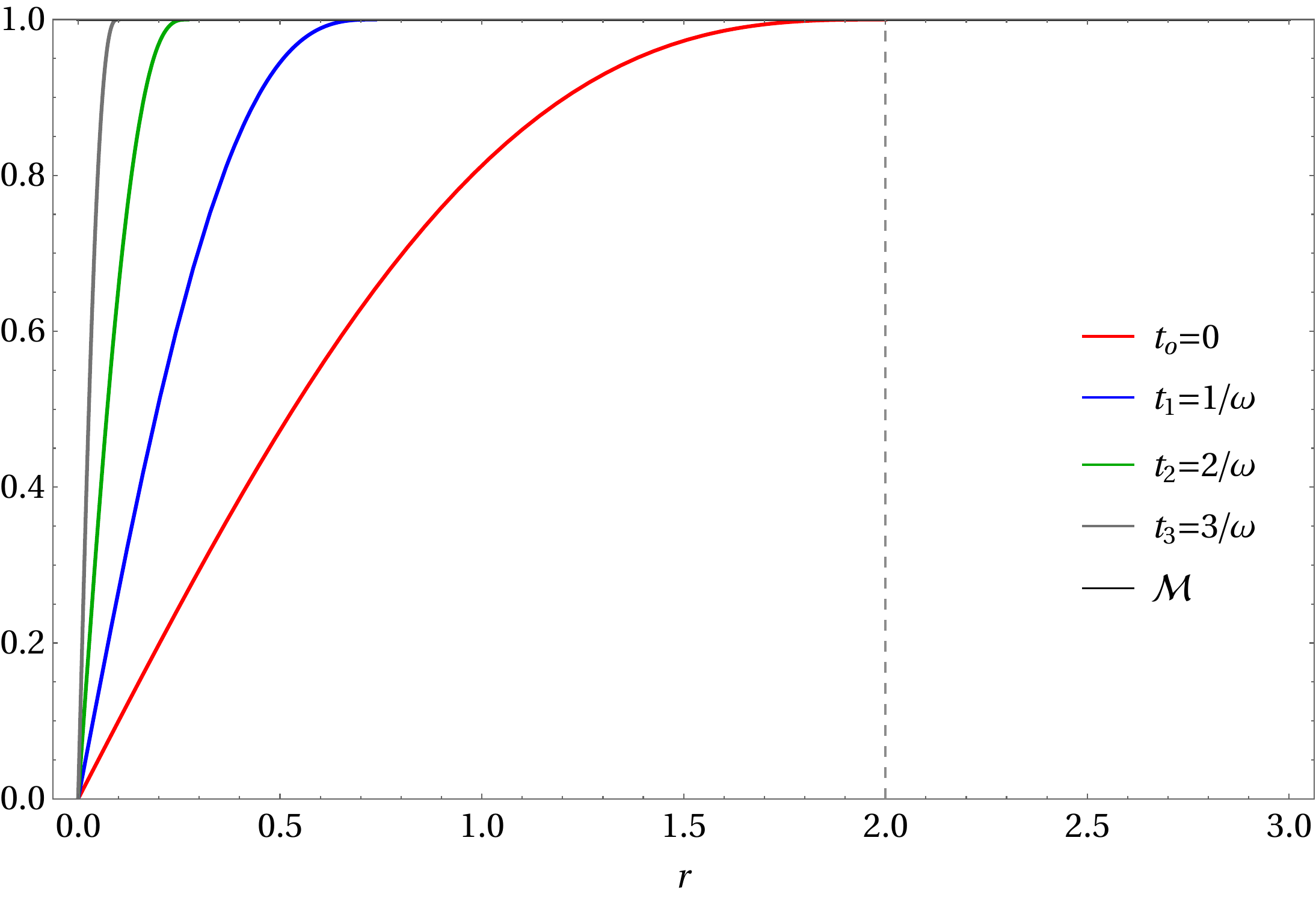}
	\caption{Mass function $m(r,t)$ in Eq.~\eqref{eq:mrt}  as a function of $r$ at $t=0$ (red), $t=1/\omega$ (blue), $t=2/\omega$ (Green), and $t=3/\omega$ (gray).
	While $m(r,t)$ approaches the profile of the constant $m(r)={\cal M}=h/2$ as $t$ increases, the matter never reaches at $r=0$ for all $t$.	
	We take $h=2.0$ and $\omega=0.1$ in this plot. This figure is taken from Ref.~\cite{Aoki:2024dyr}.} 
	\label{fig4}
\end{figure}
Fig.~\ref{fig4}  shows profiles of $m(r,t)$ as a function of $r$ at $t=0, 1/\omega, 2/\omega,  3/\omega$.  We take $h=2.0$ and $\omega=0.1$ in this figure.
A vacuum region insider the horizon develops during the collapse as $ h e^{-\omega t} \le r \le h$. 

The corresponding EMT for this metric is calculated through the Einstein equation, and one can explicitly confirm that
\beqa
T^\mu{}_\nu \nabla_\mu \xi_T^\nu = -T^\mu{}_\nu \Gamma^\nu_{\mu 0} = 0.
\eeqa
From the explicit form,
\beqa
T^0{}_0 = -{2 \partial_r m(r,t)\over 8\pi G_N r^2} ,
\eeqa
the total energy is calculated as
\beqa
E = -\int d\Sigma_\mu T^\mu{}_\nu\xi^\nu_T = -\int d\Omega\, dr\, r^2\, T^0{}_0 ={1\over G_N}\int_0^\infty dr\,\p_r m(r,t) =\frac{\cal M}{G_N}.
\label{eq:energyGC}
\eeqa
Thus the total energy $E$ is conserved and agree with the energy of the BH, which is the final state of the collapse at $t\to\infty$. 

In the collapse, Kretschmann scalar is given as
\beqa
R_{\mu\nu\alpha\beta} R^{\mu\nu\alpha\beta} =
\left\{
\begin{array}{ll}
 \dfrac{16 e^{2\omega t}}{r^4} +\cdots, & r \le h e^{-\omega t}  ,    \\
 \\
 \dfrac{12 h^2}{r^6}, &  r >  h e^{-\omega t} ,  \\
\end{array}
\right.
\eeqa
which shows no Schwarzschild singularity as $r^{-6}$ at $r=0$ develops during the collapse in the finite time $t$.

As far as the energy condition\cite{Maeda:2018hqu,Maeda:2022vld} is concerned,
the Null Energy Condition(NEC) and Strong Energy Condition (SEC) are satisfied for $\alpha :=\dfrac{\omega}{h} \le 3.13422\dots$, while the Weak Energy Condition (WEC) is satisfied for $\alpha\le 1$.
Thus the matter describing this collapse is physically acceptable. 

The final state of the collapse  is described by the mass function in the $t\to\infty$ limit as
\beqa
\lim_{t\to\infty} m(r,t) = {\cal M} \theta(r), \quad \theta(r>0)=1, \ \theta(0)=0,
\eeqa
which is the Schwarzschild vacuum except $r=0$.
This metric is the ``regularized'' Schwarzschild metric as a distribution, which leads to the non-zero EMT at the origin as\cite{Aoki:2020prb}
\beqa
T^0{}_0=T^r{}_r = -\frac{\cal M}{4\pi G_N}{\delta(r)\over r^2}, \quad
T^\theta{}_\theta=T^\phi{}_\phi = - \frac{\cal M}{8\pi G_N}{\p_r \delta(r)\over r}.
\eeqa
This EMT agrees with the one obtained  by the distributional approach\cite{Balasin:1993fn}, and
corresponds to the non-perfect fluid.
This suggests that the model of the gravitational collapse proposed by Oppenheimer-Snyder\cite{Oppenheimer:1939ue}
does not smoothly lead to the Schwarzschild BH,
since their EMT is  the dust of the contracting FLRW Universe and thus  the perfect fluid.

The energy of the BH can be calculated form the EMT as
\beqa
E_{\rm BH} &=&-\int d\Omega\, dr\, r^2 T^0{}_0 = \frac{\cal M}{G_N}\int dr\, \delta(r) = \frac{\cal M}{G_N},
\eeqa
which agrees with the energy of the collapsing star in \eqref{eq:energyGC}. Thus $t\to\infty$ limits in both metric/EMT and the energy are consistent.
This indicates that the Schwarzschild BH is made of the point-like matter at the origin.

In summary, the mass function $m(r,t)$ describes a physical process of the gravitational collapse leading to the Scwarzschild BH regularized with the $\theta$ function at $r=0$.    
 
 \section{Do gravitational fields carry energies ? -- Noether's 2nd theorem and general relativity --}
 One may think, like external electromagnetic fields, that we can assign dynamical energies to time-dependent gravitational fields,
 in order to recover a conservation of the total energy in the whole system. Namely, a loss or gain of the matter energy is compensated by a change of energy for dynamical gravitational fields, to keep the total energy intact.
 Since the time translation is a part of the general coordinate transformation of general relativity,  at first sight, it seems possible to define the conserved energy
 as a Noether charge of this symmetry.   
 Unfortunately, however, Noether's 2nd theorem invalidates such a naive thought, since the general coordinate transformation is a gauge symmetry\cite{Noether:1918zz}, and therefore we cannot assign the dynamical energy  to gravitational fields. 
 
 \subsection{Noether's 2nd theorem}
 The invariance of the Einstein-Hilbert action under the general coordinate transformation $x^\mu \to x^\mu +\xi^\mu(x)$
 implies that a current
 $
 K^\mu[\xi] = 
 \nabla_\nu \left[ \nabla^{[\mu} \xi^{\nu]}\right]
 $
 is covariantly conserved, $\nabla_\mu K^\mu =0$, without using equations of motion (EOM)\cite{Aoki:2022gez,Aoki:2022ugd},  as you can explicitly check it using the antisymmetry:
 \beqa
 \nabla_\mu K^\mu = R_{\mu\nu}{}^\mu{}_\alpha  \left[ \nabla^{[\alpha} \xi^{\nu]}\right]=R_{\nu\alpha}  \left[ \nabla^{[\alpha} \xi^{\nu]}\right] =0.
 \eeqa
 Since the conservation is not dynamical ({\it i.e.} it is conserved without using the EOM), the charge defined as
 $\int d\Sigma_\mu K^\mu$  cannot be an energy we usually use in the flat spacetime. This is the claim by Noether's 2nd theorem  in Ref.~\cite{Noether:1918zz}.

Unfortunately, Noether's 2nd theorem has not been correctly addressed in the community working on general relativity  
except some.\footnote{For example,  a paper by Utiyama~Ref.~\cite{Utiyama:1984bc} and his Japanese textbook.
In the context of scientific philosophy, see Ref.~\cite{DeHaro:2021gdv}.}
Instead of showing that the energy is conserved, an opposite argument that something conserved must be the energy seemed to be employed. 
In the major textbooks, there are mainly two types of definition of the ``energy'' in general relativity, one is the Einstein's pseudotensor, the other is the quasilocal energy, both of which are related to $K^\mu$ by some total derivative term, and are trivially conserved without using the EOM.\footnote{
While the EOM is used to rewrite the part of the identity in some cases, the fact that the conservation is merely the identity still remains.}
Thus they belong to the current of Noether's 2nd theorem. Roughly speaking, the pseudotensor corresponds to the constant $\xi^\mu=\xi^\mu_T$ for $K^\mu$ while the quasilocal energy to the covariant (non-constant) $\xi^\mu$ for $K^\mu$.
As already mentioned, since the energy conservation for these definitions is non-dynamical, it may be better to say the ``energy constraint'' instead of the energy conservation. 

Besides non-dynamical nature of these energy constraints, there are other flaws more or less related to Noether's 2nd theorem.
Firstly, these types of energies (pseudotensor and quasilocal energy) can be easily changed by adding total divergent terms or employing different $\xi^\mu$ while keeping the EOM. Thus a concept of energy in these types of definitions is very fragile.   
Secondly,  these types of energy  can be always rewritten as surface integrals.
Therefore these ``energies''  all vanish in closed space without boundaries.
Please see Ref.~\cite{Aoki:2022gez} for more issues related to Noether's 2nd theorem. 

\subsection{A brief summary: Energy loss of a binary star}  
As a brief summary of our claim, we compare two different but dynamically equivalent points of view on motions of a binary star in general relativity.

In the standard view, since a binary star is continuously loosing its energy, which is carried away by emitted gravitational waves, a period of its rotation becomes shorter and shorter.
In our view, on the other hand, 
the energy of a binary star decreases and a period of its rotation becomes shorter as a result of  Einstein equation in general relativity.
Since the matter energy is not necessarily conserved, we do not have to assign an energy to gravitational waves.
The matter energy is defined from the EMT as
\beqa
E(x^0) = \int_{\Sigma(x^0)} d\Sigma_\mu T^\mu{}_\nu \xi^\nu_T, \quad \xi^\mu_T := -\delta^\mu_0,
\eeqa     
which is covariant and local, though its value depends on a choice of the time coordinate $x^0$  and the corresponding space-like  hypersurface $\Sigma(x^0)$. 

In both views, one may ask whether 
a star or a BH can be created from a vacuum in general relativity,
where the vacuum means that $T_{\mu\nu}=0$ and $G_{\mu\nu}=0$ but $R_{\mu\nu\alpha\beta} \not= 0$ without singularity.
More explicitly we ask whether the vacuum at some time will evolve to the spacetime with $T_{\mu\nu}\not=0$ in future
according to the Einstein equation $G_{\mu\nu}+\Lambda g_{\mu\nu} = 8\pi G_N T_{\mu\nu}$ and the conservation $\nabla_\mu T^\mu{}_\nu =0$.
In the standard view, this process may occur without violating the energy conservation since the energy associated  with gravitational fields in vacuum can turn into the energy of matters.  
In our view, there is no reason to prohibit such dynamical processes since energy is no more conserved in general relativity. 
In the next section, however, we will argue that such transitions from vacuum to matter is unlikely to happen by proposing a conservation law other than energy.

\section{A conserved charge in curved spacetime}
As already discussed, the matter energy is not conserved in general in curved spacetime including general relativity, and 
Noether's 2nd theorem prevents us from including dynamical gravitational fields into the definition of the energy to recover its conservation.  
However we have pointed out that there exists a different conserved quantity associated with the EMT\cite{Aoki:2020nzm,Aoki:2023zoq,Aoki:2023ufz}.

\subsection{Geometrical setup}   
We begin with the EMT, described by
\beqa
T_{\mu\nu} &=& \varepsilon u_\mu u_\nu + P_{\mu\nu}, \quad P_{\mu\nu} u^\nu = u^\mu   P_{\mu\nu} = 0, 
\label{eq:EMT}
\eeqa
where $u^\mu$ is a time-like unit vector such that $g_{\mu\nu} u^\mu u^\nu = -1$, $\varepsilon$ is an energy density, and $P_{\mu\nu}$ is a pressure tensor. This EMT is classified as the Hawking-Ellis type I\cite{Hawking:1973uf}, and it covers standard (non-null) classical matters in $3+1$ dimensions\cite{Martin-Moruno:2018eil}. We assume that $\nabla_\mu T^\mu{}_\nu = 0$ is satisfied as usual.
 
 We prepare an initial space-like hypersurface whose intrinsic  coordinates are $y^a$ ($a=1,2,\cdots, d-1$).
 On this hypersurface, using  the EMT, we define a subset ${\cal H}_{d-1}$ as 
 \beqa
 {\cal H}_{d-1} =\{ x_P^\mu(y) \vert  \varepsilon(x_P)\not=0 \}, 
 \eeqa
and denote $H_{d-1}$ as a collection of $y$ on ${\cal H}_{d-1}$.  Thus $ y\in H_{d-1}$ implies $\varepsilon(x_P(y))\not=0$.
Note that $H_{d-1}$ can be disconnected in general.
 
 Using the time-like unit vector $u^\mu$  of the EMT,
 we define a time-like curve $x^\mu(\tau,y)$ starting from an arbitrary point $x_P(y)$ on  ${\cal H}_{d-1}$ as
 \beqa
 {d x^\mu(\tau,y)\over d\tau} &=& u^\mu(x^\mu(\tau,y)), \quad  x^\mu(0,y) = x_P(y),
 \eeqa
 where $\tau$ can be both positive and negative.
 Note that $u^\mu$ always exists on $x_P\in {\cal H}_{d-1}$ since $ \varepsilon(x_P)\not=0$.
 Collecting all curves defines a foliation of hypersurfaces ${\cal H}_{d-1}(\tau)$ as
 \beqa
 {\cal H}_{d-1}(\tau) =\{ x^\mu(\tau,y) \vert {}^\exists\tau, {}^\forall y \in H_{d-1} \}, \quad {\cal H}_{d-1}(0) ={\cal H}_{d-1}.
 \eeqa
 
 At  ${}^\forall x^\mu(\tau,y) \in {\cal H}_{d-1}(\tau)$, there exists a vector $n^\mu(\tau,y)$ normal to $ {\cal H}_{d-1}(\tau)$, which is in general different from $u^\mu(\tau,y)$.
 Here and hereafter, we use a notation such as $n^\mu(\tau,y)$ instead of $n^\mu ( x(\tau,y) )$ for simplicity. 
 For latter use, we calculate 
 \beqa
 K := \nabla_\mu u^\mu = \p_\tau \ln[ (-n\cdot u) \sqrt{h}], \quad h = \det h_{ab},
 \label{eq:K}
 \eeqa
 where $h_{ab}$ is an induced metric on the hypersurface ${\cal H}_{d-1}(\tau)$.
 For the derivation, see Ref.~\cite{Aoki:2023ufz}.
  
 \subsection{Our proposal for  a conserved current}
 We first construct a conserved current form the EMT as
 \beqa
 J^\mu(x) := T^\mu{}_\nu(x) \zeta(x) u^\nu = -\varepsilon(x)\zeta(x) u^\mu(x),
 \eeqa 
 where $\zeta(x)$ is a scalar function, introduced to satisfy a current conservation, $\nabla_\mu J^\mu = 0$.
 For spherical symmetric case, a similar type of the conserved current was porposed\cite{Kodama:1979vn}.
 
 The conservation of $J^\mu$ implies
 \beqa
 \nabla_\mu J^\mu = -u^\mu \p_\mu (\zeta\varepsilon) -\zeta\varepsilon K = -{d\over d\tau} (\zeta\varepsilon) -K (\zeta\varepsilon)=0,
 \eeqa
 where $K$ is defined in eq.~\eqref{eq:K}, and we use ${dx^\mu\over d\tau} \p_\mu ={d\over d\tau}$.
This equation is easily solved as
\beqa
\zeta \varepsilon (\tau,y) = \zeta \varepsilon (0,y)\exp\left[- \int_0^\tau d\eta\, K(\eta,y)\right] 
= \zeta \varepsilon (0,y) { (n\cdot u)\sqrt{h} (0,y)\over (n\cdot u)\sqrt{h} (\tau,y)},
\eeqa
where we use the expression of $K$ in \eqref{eq:K}. Using this the conserved current is given by
\beqa
J^\mu(\tau,y) = -\zeta(0,y)\varepsilon(0,y) n(0,y)\cdot u(0,y) \sqrt{h(0,y)}{ u^\mu(\tau,y)\over n(\tau,y)\cdot u(\tau,y) \sqrt{h(\tau,y)}},
\eeqa
which can be regarded as a current of Noether's 1st theorem\cite{Aoki:2022ugd}.

\subsection{Conserved charges}
Depending on a choice of the initial condition $\zeta(0,y)$, 
there exist many different conserved currents.
We interpret this as a conservation of each streamline defined by a tangent vector $u^\mu(\tau,y)$,  starting from a point $x^\mu(0,y)$
on a hypersurface ${\cal H}_{d-1}$.   
Indeed for a system of massive point particles interacting with each other only through gravitational interactions, the corresponding conserved current
reads\cite{Aoki:2023zoq}
\beqa
J^\mu(x) &=& \sum_{n=1}^N { \zeta_n m_n\over \sqrt{-g(x)}} \int d\tau_n\, u^\mu_n(\tau_n) \delta^{(4)}(x-x_n(\tau_n) ),
\eeqa
where $x_n(\tau_n)$ and $u^\mu_n(\tau_n)$ are  position and velocity of the $n$-th particle at its proper time $\tau_n$, respectively, $m_n$ is its mass and $\zeta_n$ is an initial condition of the scalar function $\zeta(\tau_n)$ for the  $n$-th particle. 
Our $\varepsilon(0,y)$ and $\zeta(0,y)$ correspond to $m_n$ and $\zeta_n$, respectively.
Since each particle is conserved, there are $N$ independent parameters $\zeta_n$.  

Among many different initial conditions, we take  $\zeta(0,y) \varepsilon(0,y) =1$ globaly for all $y$, which leads to
\beqa
J^\mu(\tau,y) = - {n(0,y)\cdot u(0,y) \sqrt{h(0,y)} \over n(\tau,y)\cdot u(\tau,y) \sqrt{h(\tau,y)}} u^\mu(\tau,y),
\eeqa
where the information on the initial energy density $\varepsilon(0,y)$ is unnecessary to define this current.
Consequently only geometrical quantities at $\tau=0$ are required for  $J^\mu$, and
we thus call this current the geometric conserved current. 
Our initial condition, $\zeta(0,y) \varepsilon(0,y) =1$, correspond to $\zeta_n m_n=1$ in the point particle system.

The corresponding geometrical conserved charge is given by
\beqa
Q = \int_{H_{d-1}}\, d^{d-1}y \, \sqrt{h(0,y)} \, \left[-n(0,y) \cdot u(0,y) \right]  
\eeqa
which is covariant  under the coordinate transformation of $y^a$, where we use
\beqa
\int_{{\cal H}_{d-1}} d\Sigma_\mu\cdots = \int_{H_{d-1}} d^{d-1} y \sqrt{h(\tau,y)} n_\mu(\tau, y)\cdots .
\eeqa
While $Q$ becomes  a number of the total particles as $Q=N$ in the point particle system, 
$Q$ in the above is a total ``number'' of streamlines in the fluid described by the EMT.
 The expression of $Q$ in the above is an integral of a component of $u(0,y)$ tangent to $n^\mu(0,y)$ over the hypersurface ${\cal H}_{d-1}$,
 which counts a net streamlines going through the hypersurface vertically.
 Note that $ -n(0,y) \cdot u(0,y) > 0$.

At first sight, an existence of $J^\mu$ and $Q$ looks trivial,
since it simply means the conservation of total streamlines defined by $u^\mu$.
What is non-trivial here is that  $u^\mu$ never ends or emerges.
This property seems a consequence of $\nabla_\mu T^\mu{}_\nu=0$, \footnote{This statement seems plausible but we cannot rigorously prove it so far.} 
covariant conservation of the EMT, which holds only after the EOM for matters is applied.
The covariant conservation of the EMT is also required for the Einstein equation ($G_{\mu\nu} + \Lambda g_{\mu\nu} = 8\pi G_N T_{\mu\nu}$) to be consistent, since $\nabla_\mu G^\mu{}_\nu = 0$ always hold as  the Bianchi identity. 
Therefore, the seemingly trivial conservation law is tied to an intrinsic property of  general relativity.

As an alternative choice of the global initial condition, we may take $\zeta(0,y)=1$ for all $y$, which correspond to $\zeta_n=1$ in the point particle system. Then the conserved charge $Q'$ for point particles becomes
\beqa
Q' &=& \int d^{d-1}x\, \sqrt{-g} J^0 = \sum_{n=1}^N m_n \int d\tau_n  u^0_n(\tau_n) \delta(x^0-x_n^0(\tau_n) )=\sum_{n=1}^N m_n,
\eeqa  
which is the total mass of the  point particle system. Thus
\beqa
Q' = \int_{H_{d-1}}\, d^{d-1}y \, \varepsilon(0,y) \sqrt{h(0,y)} \, \left[-n(0,y) \cdot u(0,y) \right]
\eeqa
is the total gravitational mass (not energy) of matters described by the EMT in \eqref{eq:EMT}, which is always conserved even if the total energy is not conserved during a dynamical process. 
We thus call $Q'$ a ``gravitational charge'' in general relativity, like the total charge of the electromagnetic interaction.

The existence of conserved charges discussed in this section is a reason why the vacuum in general relativity cannot transit to a non-vacuum with non-zero EMT in \eqref{eq:EMT}.  While $Q=0$ (or $Q'=0$) in the vacuum state, non-zero EMT leads to $Q\not=0$ (or $Q'\not=0$).
Therefore, the conservation of $Q$ or $Q'$ prohibits  any transitions from zero EMT to non-zero EMT, and vice versa. 

In the above discussion, we assume that densities of the conserved charges are defined to be non-negative everywhere by appropriate choices of the initial condition $\zeta(0,y)$, and $Q$ or $Q'$ is expected to satisfies this condition.
Physically this seems reasonable since the gravitational charge is expected to be positive ({\it i.e.} no negative mass).  
However, if a negative density  were locally allowed, a pair creation of positive and negative densities could become possible keeping the zero total charge condition. 
Indeed  a vacuum solution to Einstein equation except singularities was proposed in Ref.~\cite{Aoki:2022ysm}
to describe a spacetime created from and annihilated into nothing.
In the universe described by this solution, the total charge is always zero, so that the universe can emerge and vanish, while 
non-zero EMT exists at singularities but violates null, weak, strong and dominant energy conditions.  

\section{Discussions}
In this report, we have discussed the matter energy non-conservation in curved spacetime, in particular, in general relativity.
We have argued that non-conservation cannot be fixed by considering gravitational energies due to Noether's 2nd theorem for local gauge symmetries 
(the general coordinate transformation in the current case), which only leads to ``energy constraint'' as identities instead of dynamical energy conservation as a consequence of EOM.  

If one accepts the matter energy non-conservation,  one may wonder whether matter ({\it  i.e.} non-zero EMT) can emerge from  a vacuum (zero EMT)
and vice versa in general relativity.
We have answered this question negatively, by showing an existence of conserved charges $Q$ (geometrical charge) or $Q^\prime$ (gravitational charge) in a curved spacetime as long as these charge densities are always locally non-negative. 
 
Before closing this report,  we would like to mention a physical meaning of $Q$ in  special cases.
In Ref.~\cite{Aoki:2023ufz}, we have shown that $Q$ corresponds to entropy for  perfect fluids.
Thus the conserved  current is written as $J^\mu(x) = s(x) u^\mu(x)$, where
the entropy density $s$ is given by 
\beqa
s(\tau,y) =\varepsilon(\tau,x) \zeta(\tau,y) ={g(0,y)\over g(\tau,y)}, \quad g(\tau,y) :=n(\tau,y)\cdot u(\tau,y) \sqrt{h(\tau,y)} .
\eeqa 
 In the case of radiations with an equation of state $P(\tau,y)=\omega(y)\varepsilon(\tau,y)$,
 we can derive the Stefan-Boltzmann-law as
 \beqa
 \varepsilon(\tau,y) =\varepsilon(0,y)\left( {T(\tau,y)\over T(0,y)}\right)^{1+{1\over \omega(y)}}, \quad
 T(\tau,y) = T(0,y) \left({g(0,y)\over g(\tau,y)}\right)^{\omega(y)}.
 \eeqa 
 where $T(\tau,y)$ is the (local) temperature of the spacetime. 
 In 4-dimensions, uniform radiations have $\omega(y) =1/3$, which indeed leads to $\varepsilon \propto T^4$.

\section*{Acknowledgments}
I have enjoyed fruitful collaborations with Drs.~Yoshimasa Hidaka,  Kiyoharu Kawana,  Tetsuya Onogi, Kengo Shimada, Tatsuya Yamaoka and Shuichi Yokoyama.\\
I have greatly benefited from the fact that  Prof. K.K. Phua started IJMPA and MPLA 40 years ago,
where several of our papers mentioned in this report have appeared. 
I  would like to sincerely celebrate the 40th anniversary of both IJMPA and MPLA.

\end{document}